\documentclass[prc,twocolumn,showpacs,preprintnumbers,amsmath,amssymb]{revtex4-1}
\usepackage[dvips]{graphicx}
\usepackage{dcolumn}
\usepackage{bm}

\begin{document}
\title{Truncation scheme of time-dependent density-matrix approach II}
\author{Mitsuru Tohyama$^{1},$Peter Schuck$^{2,3}$}
\affiliation{$^1$Kyorin University School of Medicine, Mitaka, Tokyo
  181-8611, Japan     }
\affiliation{$^2$Institut de Physique Nucl$\acute{e}$aire, IN2P3-CNRS,
Universit$\acute{e}$ Paris-Sud, F-91406 Orsay Cedex, France}
\affiliation{$^3$Laboratoire de Physique et de Mod\'elisation des Milieux Condens\'es, CNRS and Universit\'e Joseph Fourier, 25 Av. des Martyrs, BP 166, F-38042 }

\begin{abstract}
A truncation scheme of the Bogoliubov-Born-Green-Kirkwood-Yvon hierarchy
for reduced density matrices, where a three-body density matrix is 
approximated by two-body density matrices, is improved to take into account a normalization effect.
The truncation scheme is tested for the Lipkin model. It is shown that
the obtained results are in good agreement with the exact solutions.
\end{abstract}
\maketitle
The equations of motion for reduced density matrices have a coupling scheme known as 
the Bogoliubov-Born-Green-Kirkwood-Yvon (BBGKY) hierarchy where an $n$-body density matrix couples to
$n$-body and $n+1$-body density matrices. 
To solve the equations of motion for the one-body and two-body density matrices, we need to truncate the BBGKY hierarchy
at a two-body level. The simplest truncation scheme is to approximate  
a three-body density matrix with the antisymmetrized products of the one-body and two-body density matrices neglecting the correlated
part of the three-body density matrix \cite{WC,GT}. In some cases this truncation scheme 
overestimates ground-state correlations \cite{TTS} and causes instabilities of the obtained solutions \cite{toh12} 
for strongly interacting cases. Obviously the problems originate in the
truncation scheme where the three-body correlation matrix is completely neglected \cite{schmitt,gherega}.
We have proposed a truncation scheme where the three-body correlation matrix is approximated by the products of the two-body correlation matrices.
It has been shown that the truncation scheme can remedy difficulties of the simplest truncation scheme \cite{ts2014}. However, it has been pointed out that
discrepancy between the TDDM results and the exact solutions increases with increasing interaction strength.
In this paper we show that inclusion of a normalization effect much improves agreement with the exact solutions.

We consider a system of $N$ fermions and assume that the Hamiltonian $H$ consisting of a one-body part and a two-body interaction
\begin{eqnarray}
H=\sum_\alpha\epsilon_\alpha a^\dag_\alpha a_\alpha
+\frac{1}{2}\sum_{\alpha\beta\alpha'\beta'}\langle\alpha\beta|v|\alpha'\beta'\rangle
a^\dag_{\alpha}a^\dag_\beta a_{\beta'}a_{\alpha'},
\label{totalH}
\end{eqnarray}
where $a^\dag_\alpha$ and $a_\alpha$ are the creation and annihilation operators of a particle at
a single-particle state $\alpha$.
TDDM gives
the coupled equations of motion for the one-body density matrix (the occupation matrix) $n_{\alpha\alpha'}$
and the two-body density matrix $\rho_{\alpha\beta\alpha'\beta'}$.
These matrices are defined as
\begin{eqnarray}
n_{\alpha\alpha'}(t)&=&\langle\Phi(t)|a^\dag_{\alpha'} a_\alpha|\Phi(t)\rangle,
\\
\rho_{\alpha\beta\alpha'\beta'}(t)&=&\langle\Phi(t)|a^\dag_{\alpha'}a^\dag_{\beta'}
 a_{\beta}a_{\alpha}|\Phi(t)\rangle,
 \label{rho2}
\end{eqnarray}
where $|\Phi(t)\rangle$ is the time-dependent total wavefunction

\noindent
$|\Phi(t)\rangle=\exp[-iHt/\hbar] |\Phi(t=0)\rangle$.
The equations in TDDM are written as
\begin{eqnarray}
i \hbar\dot{n}_{\alpha\alpha'}&=&
(\epsilon_{\alpha}-\epsilon_{\alpha'}){n}_{\alpha\alpha'}
\nonumber \\
&+&\sum_{\lambda_1\lambda_2\lambda_3}
[\langle\alpha\lambda_1|v|\lambda_2\lambda_3\rangle \rho_{\lambda_2\lambda_3\alpha'\lambda_1}
\nonumber \\
&-&\rho_{\alpha\lambda_1\lambda_2\lambda_3}\langle\lambda_2\lambda_3|v|\alpha'\lambda_1\rangle],
\label{n2}
\end{eqnarray}
\begin{eqnarray}
i\hbar\dot{\rho}_{\alpha\beta\alpha'\beta'}&=&
(\epsilon_{\alpha}
+\epsilon_{\beta}
-\epsilon_{\alpha'}
-\epsilon_{\beta'}){\rho}_{\alpha\beta\alpha'\beta'}
\nonumber \\
&+&\sum_{\lambda_1\lambda_2}[
\langle\alpha\beta|v|\lambda_1\lambda_2\rangle\rho_{\lambda_1\lambda_2\alpha'\beta'}
\nonumber \\
&-&\langle\lambda_1\lambda_2|v|\alpha'\beta'\rangle\rho_{\alpha\beta\lambda_1\lambda_2}]
\nonumber \\
&+&\sum_{\lambda_1\lambda_2\lambda_3}
[\langle\alpha\lambda_1|v|\lambda_2\lambda_3\rangle\rho_{\lambda_2\lambda_3\beta\alpha'\lambda_1\beta'}
\nonumber \\
&+&\langle\lambda_1\beta|v|\lambda_2\lambda_3\rangle\rho_{\lambda_2\lambda_3\alpha\alpha'\lambda_1\beta'}
\nonumber \\
&-&\langle\lambda_1\lambda_2|v|\alpha'\lambda_3\rangle\rho_{\alpha\lambda_3\beta\lambda_1\lambda_2\beta'}
\nonumber \\
&-&\langle\lambda_1\lambda_2|v|\lambda_3\beta'\rangle\rho_{\alpha\lambda_3\beta\lambda_1\lambda_2\alpha'}],
\label{N3C2}
\end{eqnarray} 
where $\rho_{\alpha\beta\gamma\alpha'\beta'\gamma'}$ is a three-body density-matrix.
In Refs. \cite{WC,GT} the BBGKY hierarchy is truncated by replacing the three-body density matrix with
the antisymmetrized product of $n_{\alpha\alpha'}$ and $\rho_{\alpha\beta\alpha'\beta'}$ neglecting the correlated part $C_{\alpha\beta\gamma\alpha'\beta'\gamma'}$
of $\rho_{\alpha\beta\gamma\alpha'\beta'\gamma'}$.
Our previous truncation scheme for Eq. (\ref{N3C2}) is the following \cite{ts2014}: 
Instead of neglecting $C_{\alpha\beta\gamma\alpha'\beta'\gamma'}$ 
we use 
\begin{eqnarray}
C_{\rm p_1p_2h_1p_3p_4h_2}&=&\sum_{\rm h}C_{\rm hh_1p_3p_4}C_{\rm p_1p_2h_2h},
\label{purt1}\\
C_{\rm p_1h_1h_2p_2h_3h_4}&=&\sum_{\rm p}C_{\rm h_1h_2p_2p}C_{\rm p_1ph_3h_4},
\label{purt2}
\end{eqnarray}
where ${\rm p}$ and ${\rm h}$ refer to particle and hole states, respectively.
These expressions were derived from perturbative consideration \cite{ts2014} using
the following CCD (Coupled-Cluster-Doubles)-like ground state wavefunction $|Z\rangle$ \cite{shavitt}
\begin{eqnarray}
|Z\rangle=e^Z|{\rm HF}\rangle
\end{eqnarray}
with 
\begin{eqnarray}
Z=\frac{1}{4}\sum_{\rm pp'hh'}z_{\rm pp'hh'}a^\dag_{\rm p}a^\dag_{\rm p'}a_{\rm h'}a_{\rm h},
\end{eqnarray}
where $|{\rm HF}\rangle$ is the HF ground state and  
$z_{\rm pp'hh'}$ is antisymmetric under the exchanges of ${\rm p} \leftrightarrow {\rm p'}$ and ${\rm h} \leftrightarrow {\rm h'}$.
Assuming that $z_{\rm pp'hh'}$ is small, that is, $|Z\rangle\approx (1+Z)|{\rm HF}\rangle$ 
and $C_{\rm pp'hh'}\approx z_{\rm pp'hh'}$, where $C_{\rm pp'hh'}$ is the correlated part of $\rho_{\rm pp'hh'}$, we arrived at Eqs. (\ref{purt1}) and (\ref{purt2}).
It has been pointed out \cite{ts2014} in the applications to model Hamiltonians that 
in strongly interacting regions where perturbative treatment is not justified the truncation scheme of Eqs. (\ref{purt1}) and (\ref{purt2}) 
underestimates correlation effects. This indicates that the coupling to higher-order reduced density matrices plays a role in reducing the three-body correlation matrix.
Our new truncation scheme is to include such a reduction effect using the normalization $\langle Z|Z\rangle$ of the total wavefunction.
Assuming that the three-body correlation matrix is calculated using the wavefunction $|Z\rangle=(1+Z)|{\rm HF}\rangle$, which gives   
the normalization
\begin{eqnarray}
{\cal N}&=&\langle Z|Z\rangle=1+\frac{1}{4}\sum_{\rm pp'hh'}z_{\rm pp'hh'}z^*_{\rm pp'hh'}
\label{norm}
\end{eqnarray}
we express the three-body correlation matrix as
\begin{eqnarray}
C_{\rm p_1p_2h_1p_3p_4h_2}&=&\frac{1}{\cal N}\sum_{\rm h}z^*_{\rm p_3p_4hh_1}z_{\rm p_1p_2h_2h},
\label{purt11}\\
C_{\rm p_1h_1h_2p_2h_3h_4}&=&\frac{1}{\cal N}\sum_{\rm p}z^*_{\rm p_2ph_1h_2}z_{\rm p_1ph_3h_4}.
\label{purt22}
\end{eqnarray}
When Eqs. (\ref{purt11}) and (\ref{purt22}) are evaluated, we approximate $z_{\rm pp'hh'}$ and $z^*_{\rm pp'hh'}$ by $C_{\rm pp'hh'}$
and $C_{\rm hh'pp'}$, respectively.
We refer to this truncation scheme as TDDM and the truncation scheme of Eqs. (\ref{purt1}) and (\ref{purt2}) as TDDM$_0$, respectively.
The normalization ${\cal N}$ thus introduced plays a role in reducing the three-body correlation matrix.
The reader may be somewhat puzzled by this procedure in view of Eq. (A5) in \cite{Tohy-prb93} where no norm appears. However, as we will see with the applications, neglecting simply the four-body correlation matrix is not such a good approximation in the strong coupling regime. We, therefore, were guided by Eqs. (\ref{purt11}) and (\ref{purt22}) to introduce also a norm into Eqs. (\ref{purt1}) and (\ref{purt2}). This is a slightly ad hoc procedure but, as we will see, this very much improves the results.

We test TDDM for the Lipkin model.
The Lipkin model \cite{Lip} describes an $N$-fermions system with two
$N$-fold degenerate levels with energies $\epsilon/2$ and $-\epsilon/2$,
respectively. The upper and lower levels are labeled by quantum number
$p$ and $-p$, respectively, with $p=1,2,...,N$. We consider
the standard Hamiltonian
\begin{equation}
{H}=\epsilon {J}_{z}+\frac{V}{2}({J}_+^2+{J}_-^2),
\label{elipkin}
\end{equation}
where the operators are given as
\begin{eqnarray}
{J}_z&=&\frac{1}{2}\sum_{p=1}^N(a_p^{\dag}a_p-{a_{-p}}^{\dag}a_{-p}), \\
{J}_{+}&=&{J}_{-}^{\dag}=\sum_{p=1}^N a_p^{\dag}a_{-p}.
\end{eqnarray}

\begin{figure}
\resizebox{0.5\textwidth}{!}{%
\includegraphics{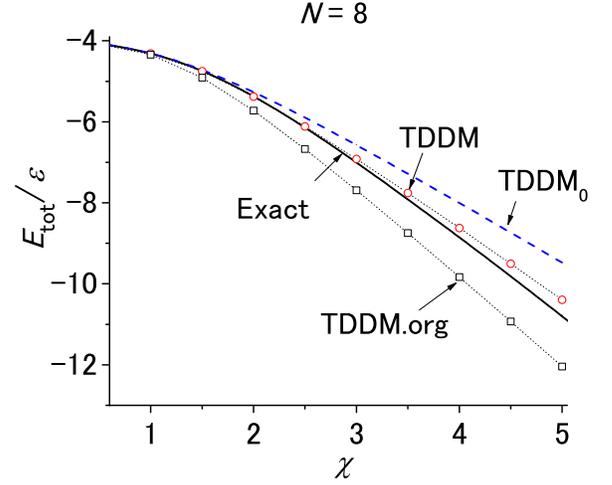}
}
\caption{Ground-state energy in TDDM (circles) 
as a function of $\chi=|V|(N-1)/\epsilon$ for $N=8$.
The dashed line depicts the results in TDDM$_0$ where the three-body correlation matrix is given by eqs. (\ref{purt1}) and (\ref{purt2}).
The squares show the results in the original truncation scheme where the three-body correlation matrix is neglected.
The exact values are given by the solid line.}
\label{E8}
\end{figure}
\begin{figure}[h]
\resizebox{0.5\textwidth}{!}{%
\includegraphics{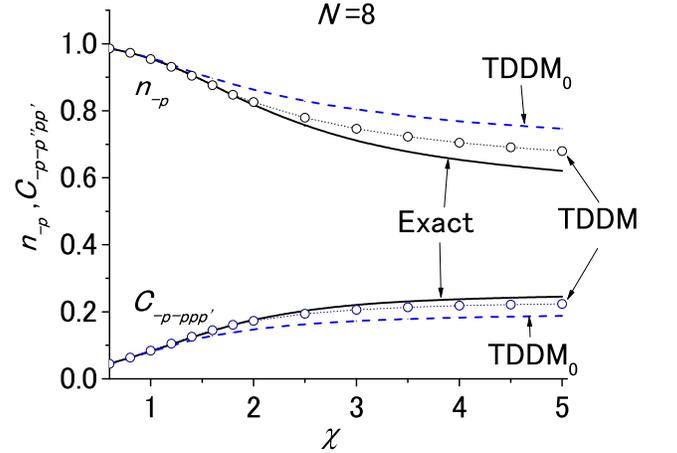}
}
\caption{Same as Fig. \ref{E8} but for the occupation probability $n_{-p}$ of the lower state and the 2p-2h element $C_{-p-p'pp'}$ of the two-body correlation matrix.
.}
\label{8nC}
\end{figure}
\begin{figure}[h]
\resizebox{0.5\textwidth}{!}{%
\includegraphics{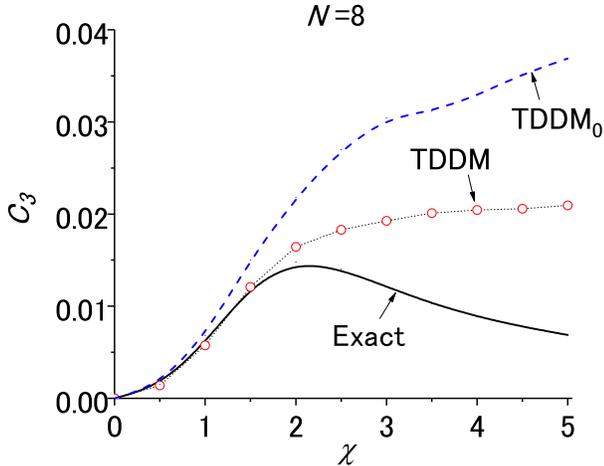}
}
\caption{Same as Fig. \ref{E8} but for the three-body correlation matrix $C_{-pp'p''p-p'p''}$.}
\label{8c3}
\end{figure}

The ground-state energy calculated in TDDM (open circles) is shown in Fig. \ref{E8} as a function of $\chi=|V|(N-1)/\epsilon$ for $N=8$.
The results in TDDM$_0$ and the exact values are given with the dashed and solid lines, respectively.
The results in the original truncation scheme (TDDM.org) where the three-body correlation matrix is neglected are shown with the squares.
The results in TDDM are obtained using an adiabatic method explained in Ref. \cite{ts2014,adiabatic}.
TDDM.org overestimates the correlation effects. TDDM$_0$ cures this problem but underestimates the correlation effects in the strongly interacting region. 
The agreement with the exact solutions is much improved in TDDM. The occupation probability $n_{-p}$ and the correlation matrix $C_{-p-p'pp'}$ in TDDM 
are also closer to the exact values than those in TDDM$_0$ as shown in Fig. \ref{8nC}. The value of ${\cal N}$ at $\chi=5$ is $2.0$.
Thus the normalization factor ${\cal N}$ in Eqs. (\ref{purt11}) and (\ref{purt22}) 
plays an important role in suppressing the three-body correlation matrix with increasing interaction strength.
This is explicitly shown in Fig. \ref{8c3} where the values of $C_{-pp'p''p-p'p''}$ calculated in TDDM (circles) 
are compared with those in TDDM$_0$
(dashed line) 
and the exact values (solid line). The normalization factor drastically reduces $C_{-pp'p''p-p'p''}$ in TDDM$_0$ though TDDM cannot 
reproduce the exact values in strong coupling. 
One should realize that we are considering values of the coupling constant ($\chi > 1$) which are deeply in the symmetry broken phase. 
Actually the strong coupling limit can very well be treated in the Lipkin model by changing the single particle basis and performing a Hartree-Fock RPA calculation (for $\chi \rightarrow \infty$, this yields the exact result). The critical region for finite systems is the one around the instability point $\chi=1$. We see that the present approach gives excellant results there.
\begin{figure}
\resizebox{0.5\textwidth}{!}{%
\includegraphics{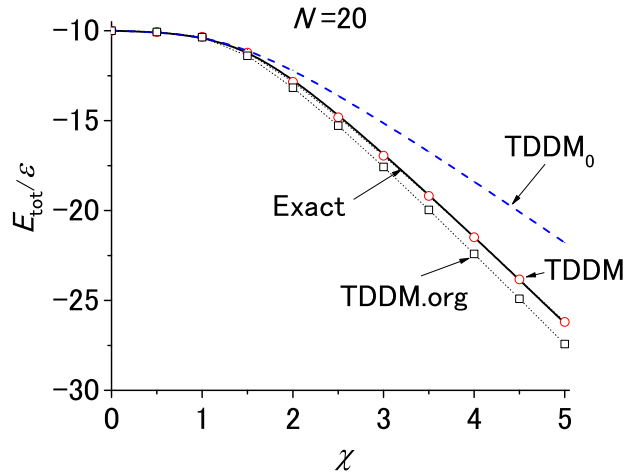}
}
\caption{Same as Fig. \ref{E8} but for $N=20$.}
\label{E20}
\end{figure}
\begin{figure}[h]
\resizebox{0.5\textwidth}{!}{%
\includegraphics{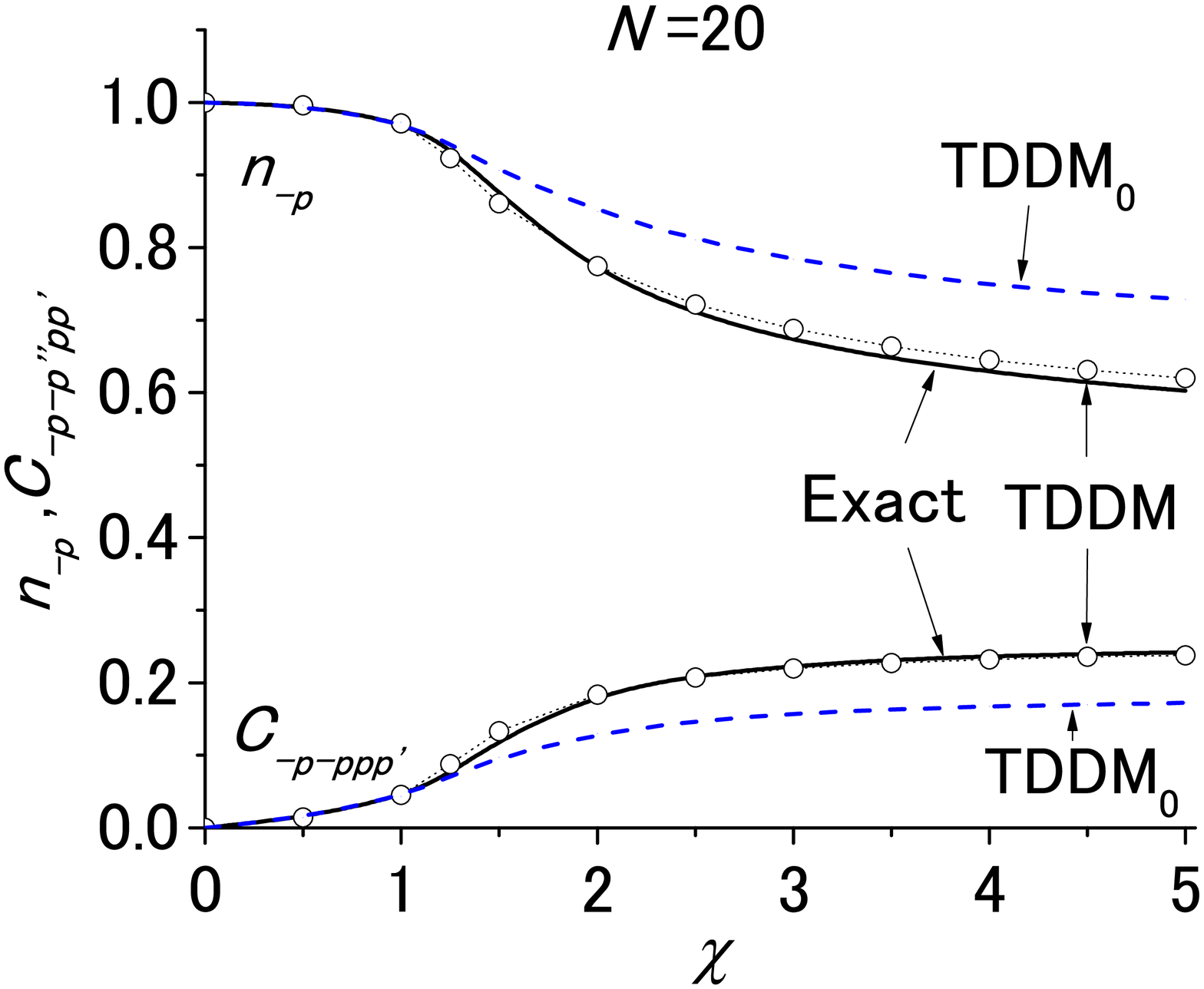}
}
\caption{Same as Fig. \ref{8nC} but for $N=20$.
}
\label{20nC}
\end{figure}
\begin{figure}[h]
\resizebox{0.5\textwidth}{!}{%
\includegraphics{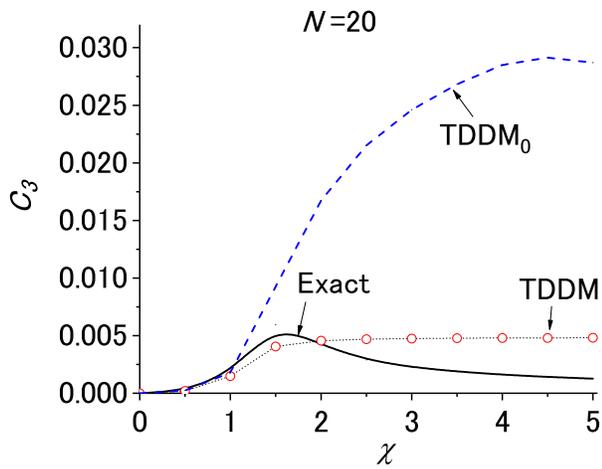}
}
\caption{Same as Fig. \ref{8c3} but for $N=20$.
.}
\label{20c3}
\end{figure}

The ground-state energies calculated in TDDM (open circles) are also shown in Fig. \ref{E20} for $N=20$.
As seen in Fig. \ref{E20},
TDDM.org becomes a good truncation scheme for $N=20$. 
The occupation probability and $C_{-p-p'pp'}$ are shown in Fig. \ref{20nC}.
The good agreement of the TDDM results with the exact solutions is also seen for the large $N$ system.
 At $\chi=5$ the value of ${\cal N}$ for $N=20$ is $6.6$.
The three-body correlation matrix is shown in Fig. \ref{20c3}.
The normalization factor drastically reduces $C_{-pp'p''p-p'p''}$ in TDDM$_0$ and the TDDM values become close to
the exact values. The value of the three-body correlation matrix for $N=20$ is much smaller than that for $N=8$. This agrees with the fact that TDDM.org which neglects
the three-body correlation matrix becomes better with increasing $N$.
Thus the importance of the three-body correlation matrix in the Lipkin model decreases with increasing number of particles. 
Let us explain this point in some more detail.
The three-body density matrix $\rho_{-pp'p''p-p'p''}$ is related to $J_+$, $J_-$ and $J_z$ as
\begin{eqnarray}
\sum_{pp'p''}\rho_{-pp'p''p-p'p''}&=&\langle\Phi_0|J_+J_-(J_z+\frac{1}{2}\hat{N})-(J_z+\frac{1}{2}\hat{N})^2
\nonumber \\
&-&J_+J_-+J_z+\frac{1}{2}\hat{N}|\Phi_0\rangle,
\label{rho3}
\end{eqnarray}
where $|\Phi_0\rangle$ is the ground-state wavefunction and $\hat{N}$ is the number operator. For large values of $N$ and $|V|$, the first term on the right-hand side of the above equation is
dominant and given by  
$\sum_{pp'p''}\rho_{-pp'p''p-p'p''}\approx {N}^3/{8}$,
where the approximation $|\Phi_0\rangle\approx|jm\rangle$ with $j=\frac{N}{2}$ and $m=0$ is used.
Here, $|jm\rangle$ is an eigenstate of ${\bm J}^2$ and $J_z$.
The left-hand side of Eq. (\ref{rho3}) is also expressed by the correlation matrices as
\begin{eqnarray}
\sum_{pp'p''}\rho_{-pp'p''p-p'p''}&=&N(N-1)(N-2)n_pC_{-pp'p-p'}
\nonumber \\
&+&N(N-1)^2C_{-pp'p''p-p'p''}
\nonumber \\
&+&N(N-1)[n_{p}C_{-ppp-p}-n_{-p}C_{pp'pp'}
\nonumber \\
&-&n_{-p}n_p^2-n_p C_{-pp'-pp'}].
\label{rho31}
\end{eqnarray}
For large values of $N$ and $|V|$, $n_p=\langle\Phi_0|J_z+\hat{N}/2|\Phi_0\rangle/N\approx 1/2$ and $C_{-pp'p-p'}=\langle\Phi_0|J_+J_-|\Phi_0\rangle/N^2\approx 1/4$, and
the last two lines can be neglected. This means that in such a limit the first term on the right-hand
side of Eq. (\ref{rho31}) becomes $N^3/8$ and consequently $C_{-pp'p''p-p'p''}\approx 0$.

Though it is not presented in this paper,
we have also applied TDDM to the one-dimensional Hubbard model and observed better agreement with the exact solutions than TDDM$_0$.
However, the improvement from TDDM$_0$ to TDDM is small because TDDM$_0$ is already a good approximation in that model.

In summary we proposed a new truncation scheme of the BBGKY hierarchy where the normalization factor of the total wavefunction is included when
the three-body correlation matrix is approximately calculated. We tested it for the ground states of the Lipkin model and obtained good agreement with
the exact solutions independently of the number of particles.
It was discussed that the normalization factor plays a role in suppressing overgrowth
of the three-body correlation matrix with increasing interaction strength.
It was also pointed out that the original truncation scheme where the three-body correlation matrix is completely neglected becomes
a better approximation with increasing number of particles.

\end{document}